\documentclass[twocolumn,showpacs,preprintnumbers,amsmath,amssymb]{revtex4}

\newcommand{\equ}[1]{(\protect\ref{#1})}
\newcommand{\bA}[1]{\hat{b}_#1}
\newcommand{\bC}[1]{\hat{b}_#1^\dagger}
\newcommand{\aA}[1]{\hat{a}_#1}
\newcommand{\aC}[1]{\hat{a}_#1^\dagger}
\newcommand{\state}[1]{\left| {#1} \right>}
\newcommand{\astate}[1]{\left< {#1} \right|}

\usepackage{graphicx}
\usepackage{dcolumn}

\usepackage{amssymb}

\usepackage{bm}
\usepackage{color}

\def\H{\mathcal H}
\def\tph{\widetilde{\phi}}
\def\tps{\widetilde{\psi}}

\def\ba{\bar{a}}
\def\bb{\bar{b}}

\begin{document}

\title{Field theory of molecular cooperators}

\author{Jordi Pi\~nero$^{1,2}\footnote{jordi.pinero@upf.edu}$ and Ricard Sol\'e$^{1,2,3}\footnote{ricard.sole@upf.edu}$}
\affiliation{$^{1}$ICREA-Complex Systems  Lab, Universitat Pompeu Fabra,   08003   Barcelona,   Spain}
\affiliation{$^{2}$Institut de Biologia Evolutiva (CSIC-UPF), Psg Maritim Barceloneta, 37, 08003 Barcelona, Spain}
\affiliation{$^{3}$Santa Fe Institute, 399 Hyde Park Road, Santa Fe NM 87501, USA}

\date{\today}

\begin{abstract}
It has been suggested that major transitions in evolution require the emergence of 
novelties, often associated to the cooperative behaviour of previously existing objects or agents. 
A key innovation involves the first cooperative interactions among molecules in a prebiotic 
biosphere. One of the simplest scenarios includes two molecular species capable of helping each other 
forming a catalytic loop or {\em hypercycle}. The second order kinetics of the hypercycle implies 
a hyperbolic growth dynamics, capable of overcoming some selection barriers associated to non-cooperative 
molecular systems. Moreover, it has been suggested that molecular replicators 
might have benefited from a limited diffusion associated to their attachment to surfaces: evolution 
and escape from extinction might have been tied to living on a surface. In this paper we propose a 
field theoretical model of the hypercycle involving reaction and diffusion through the use of a 
many-body Hamiltonian. This treatment allows a
characterisation of the spatially correlated dynamics of the system, where the critical dimension is found to be $d_c=2$. 
We discuss the role of surface dynamics as a selective advantage for the system's survival. 

\end{abstract}

\keywords{Complex systems; Reaction-Diffusion Field Theory; Hypercycle.}
\pacs{05.40.-a, 89.75.-k, 64.60.-i, 87.15.Zg}

\maketitle

\section{Introduction}

Life on our planet has experienced several key innovation events since its early 
appearance \cite{Smith97}. 
The emergence of self-replicating molecular entities pervaded the rise of complex life forms, 
including those having an embodied, protocellular organisation. Each of these so called 
{\em major transitions} incorporated some kind of novel way of enhancing cooperation among 
simpler subsystems. Most transitions are essentially tied to improving the selective advantage 
of the agents inhabiting the new complexity level, and that means exploiting some class 
of cooperative interaction at the higher scale. In particular, it has been conjectured that, in order to 
achieve higher levels of genetic information, moving beyond small, unreliable chains of molecular 
units, cooperation among different replicating molecules was required \cite{Smith97,Szathmary1997,SchusterEvo}. 

One of the most celebrated candidates to explain this class of phenomena and its 
implications is Eigen's theory of the hypercycle \cite{Eigen, Kuppers2012}
that proposes a key role of mutually enhancing replication among complementary subunits. Specifically, 
the general hypercycle involves a population of $n$ coupled molecular species described by a 
set of concentrations $\{ \rho_i \}$ with $i=1, ..., n$ whose replication is dependent upon 
the presence of another species that act as catalyst, forming a closed loop. Typically, the deterministic 
dynamics is given by a set of first-order coupled equations 

 \begin{equation}
{d \rho_i \over dt} =  \Lambda_i ({\bf \rho} ; \{ \Gamma_{ij} \} )
\end{equation}
with ${\bf \rho} = (\rho_1, ..., \rho_n )$ and 
where $ \{ \Gamma_{ij} \}$ is a set of parameters. In its original formulation, these equations included a 
reaction and an outflow term, i. e. $ \Lambda_i = \Omega_i({\bf \rho}) - \rho_i F(\rho)$. Here $\Omega_i({\bf \rho})$ includes 
the reactions among molecular species and the last term is a continuous dilution flux $F$ required to maintain a 
constant population $c=\sum_j \rho_j$ of molecules. Using $c=1$, it is not difficult to show that 
the dynamics is described by a general form 
 \begin{equation}
{d \rho_i \over dt} =  \Omega_i({\bf \rho}) - \rho_i \sum_{j=1}^n \Omega_j({\bf \rho})
\end{equation}
Which automatically satisfies the so called constant population constraint (CPC), i. e. $dc/dt=0$. 

The specific form of the original $n$-member hypercycle is a closed loop of catalytic reactions. 
This can be described by the set of equations:
 \begin{equation}
{d \rho_i \over dt} =  \Gamma_{i,i-1} \rho_i \rho_{i-1} - \rho_i \sum_{j=1}^n \Gamma_{j,j-1} \rho_j \rho_{j-1}
\end{equation}
where the $i$-th member of the cycle catalyses the growth of the $(i+1)$-th species while it is also 
helped by the $i-1$-th one in the loop (with the constraint $i+1 \rightarrow n$ when $i=n$) \cite{Eigen, Kuppers2012, Boerjlist1991, Cronhjort1994,Cronhjort1997,Chacon1995,JTBspatialhypercycles,Attolini2006}. 

This model has several remarkable properties. One is that it properly describes the logic 
of molecular cooperation networks \cite{PaulHiggs15} where cross-catalytic molecules are 
present (particularly RNA or peptide chains). Secondly, the hypercycle involves a superexponential kinetics 
that, under ideal circumstances predicts a finite-time singularity. An important implication 
of this growth dynamics is that the hypercycle is capable of overcoming exponentially growing, Darwinian 
replicators \cite{Szathmary1997}. Finally, although we will restrict our discussion to simple, molecular 
replicators, it is worth mentioning that it has also been used as a relevant descriptor of other 
out of equilibrium systems, including ecosystems \cite{Smith97,Wilkinson2006} and economic networks \cite{JohnPadget}. 
Similarly, it has been suggested that in a homogeneous (well mixed) setting this cooperative 
structure would easily break down (and become extinct) under the presence of parasites \cite{MSmith1979}.
 
The emergence of the first hypercycles is a specially relevant problem. If we consider the problem of replicating 
molecules experiencing mutation and selection, a key result of theoretical models of evolution is that 
there are sharp limits to the complexity that such system can achieve \cite{quasispecies}. For a given mutation rate, 
it can be shown that there is a limit to genome size (scaling up as the inverse of mutation rate). How can this be 
overcome? One way is to make several genomes cooperate, thus effectively moving beyond the 
so called error threshold. In an early biosphere, before 
closed compartments might have facilitated chemical reactions, pairwise interactions among the 
members of the hypercycle must have been difficult. Stochastic effects and molecular decay might counterbalance the 
potential for rapid growth predicted by the ideal model. Not surprisingly, several studies have focused on 
the role played by space in the evolution and persistence of these structures 
\cite{Boerjlist1991, Cronhjort1994,Cronhjort1997,Chacon1995,JTBspatialhypercycles,Attolini2006}.

\begin{figure}
        \begin{center}
          \includegraphics[width= 7.5 cm]{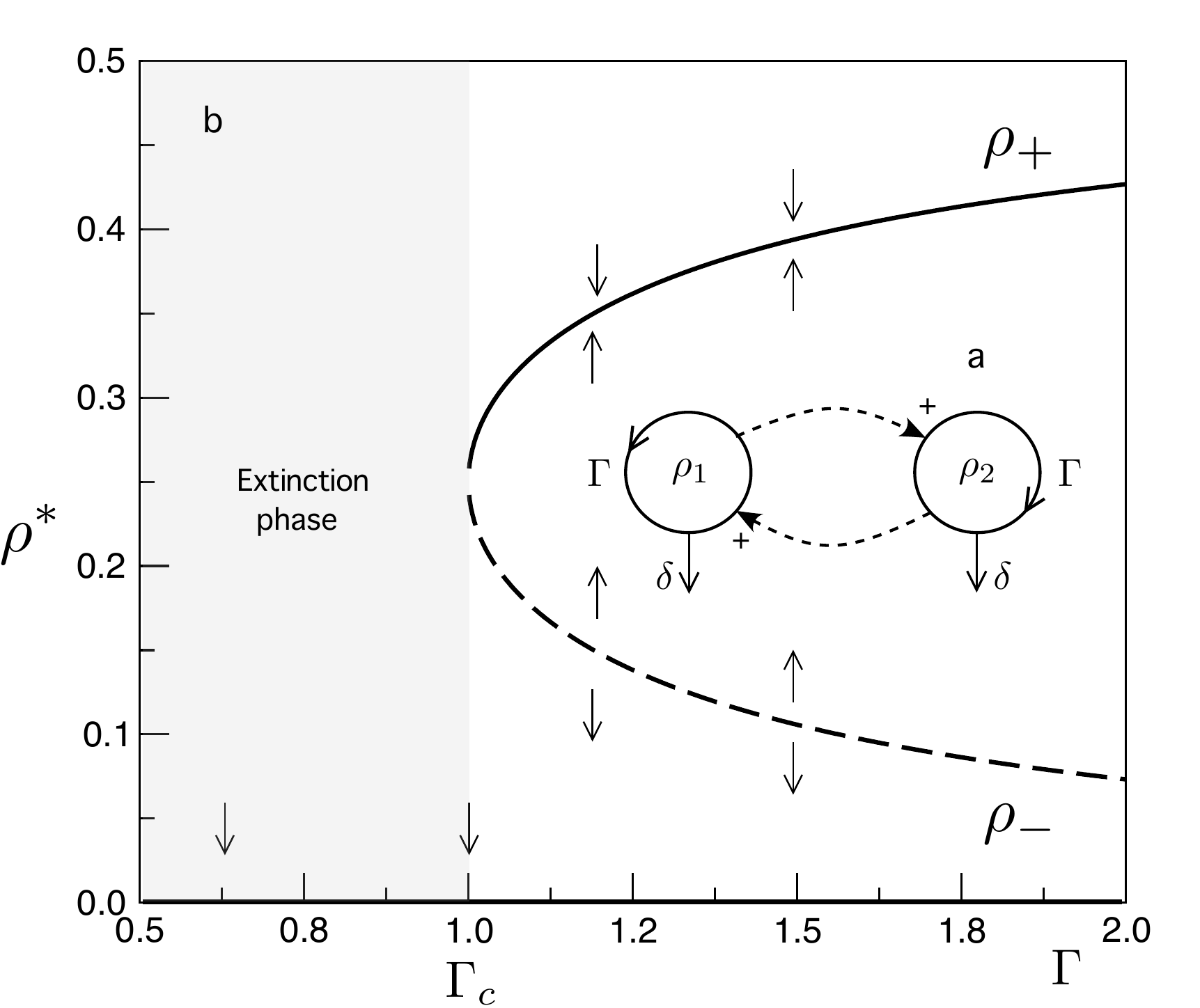}

          \caption{Bifurcations in the simple symmetric hypercycle. The logic organisation of this 
          pairwise cooperative loop is outlined in (a). In (b) the fixed points of the system are plotted against 
          the control parameter $\Gamma$ 
          (here $K=1$ and $\delta=0.125$). A first order transition is at work. For $\Gamma<\Gamma_c$ an absorbing state (extinction of both 
          cooperators) is present, whereas a stable hyper cycle can form if $\Gamma>\Gamma_c$ and the initial state 
          is such that $\rho(0)>\rho_-$.}

          \label{fig:1}
        \end{center}
      \end{figure}

The simplest model considers two classes of replicating chains \cite{JTBspatialhypercycles} which can self-replicate 
provided that a pairwise interaction takes place (see Fig.\ref{fig:1}). This requirement implies that a cross-catalytic 
reaction is needed and a second-order kinetics underlies the population dynamics of both chains. Such kind of 
reaction-diffusion dynamical system can be naturally explored 
using field theoretical approximations, which allow to understand the universal features of certain classes 
of reaction kinetics and the role played by spatial degrees of freedom. This approach has been 
successful in approaching a wide range of problems including chemical kinetics, population dynamics or 
molecular processes both within cells and in cell-free contexts \cite{fieldtheorypapers} as well as to the 
dynamics of quasispecies \cite{quasispeciesFT}. This is also the approach 
we take here to investigate some universal properties of hypercycles.

The paper is organised as follows. In section II we outline the basic results associated to the mean field 
description of a two-member hypercycle. In section III a master equation is introduced on a
discrete spatial lattice and make use of Doi-Peliti's formalism to obtain a hamiltonian description of the system.
Then the continuum limit is taken such that a field theoretical action may be written. We perform a perturbative one-loop computation around a non-trivial vacuum corresponding to the stable active state. We will finally apply renormalization group ideas in order to discuss the couplings flow equations on a coarse-graining scheme. This leads to a unique insight on the role of dimensionality for the spatial dynamics of the system. In particular, we will be able to observe these effects by plotting a phase diagram of active/inactive states on the large scale or infra-red limit. In section IV we briefly discuss potential extensions to $N$-dimensional hypercycles. In section V we summarise our basic results and discuss possible further work.

\section{Mean field theory of the symmetric hypercycle}

The simplest model of a hypercycle involves a 
two-member system with symmetric interactions, as summarised in Fig 1. 
In this section we make the assumption of a well-mixed scenario, where the population densities of each member of the 
cooperative loop, hereafter indicated as $A$ and $B$, will be given by $\rho_1$ and $\rho_2$, 
respectively. As shown in the diagram, each molecular species replicates with help from the other partner 
while both degrade. 

The hypercycle equations for this system would read, in general, as a pair of coupled cooperation-like differential equations:
\begin{eqnarray}
&&{d\rho_1 \over dt} = \Gamma_{12}\rho_1 \rho_2 \left (1 - \frac{\rho_1 +\rho_2}{K} \right ) - \delta_1 \rho_1  \label{MFT1}\\
&&{d\rho_2 \over dt} = \Gamma_{21}\rho_1 \rho_2 \left  (1 - \frac{\rho_1 +\rho_2}{K} \right ) - \delta_2 \rho_2  \ . \label{MFT2}
\end{eqnarray}
The coefficients $ \Gamma_{ij}$ stand for the replication rate of each molecular species under the presence 
of the second one. There is a limit to the maximum population of molecular replicators provided by the 
carrying capacity $K$. The last term stands for linear degradation rates. It is worth mentioning that this model includes 
several particular cases, some of which have been studied in previous papers \cite{JTBspatialhypercycles}. For example, if $K \gg 1$ and $\Gamma_{21}=0, \delta_2=0$ we recover a model by Schnerb et al \cite{SolomonPNAS2000}.

Under the symmetry assumption, $\Gamma_{12}=\Gamma_{21}\equiv\Gamma$ and $\delta_1=\delta_2\equiv\delta$, equal populations are achieved at equilibrium, so the fixed points are, $\rho^*_1 = \rho^*_2 = \rho$. A good description of the dynamics \cite{HYPalternative} 
is provided by the single differential equation model for the density $\rho$: 
\begin{eqnarray}
{d\rho \over dt} = \Gamma \rho^2 \left (1 -  \frac{2\rho}{K} \right) - \delta \rho \,.
\end{eqnarray}
The fixed points of this equation are the trivial one, $\rho_0=0$, associated to the 
extinction (absorbing) phase plus two solutions 
 \begin{equation}
\rho_{\pm} =  \frac{K}{4} \left ( 1 \pm \sqrt {1 - {\Gamma_c \over \Gamma} } \right) \,,
\end{equation}
where we use $\Gamma_c=8\delta/K$. The (linear) stability of these 
fixed points is determined by the sign of 
 \begin{equation}
 \label{eq:cpcmft}
\lambda_{\Gamma}(\rho_s) =  \left [ {d \dot{\rho} \over d\rho} \right ]_{\rho_s} ,
\end{equation}
for each $\rho_s$. Stable and unstable points are characterised by 
negative and positive values of $\lambda$, respectively.

     \begin{figure}
        \begin{center}
          \includegraphics[width= 7. cm]{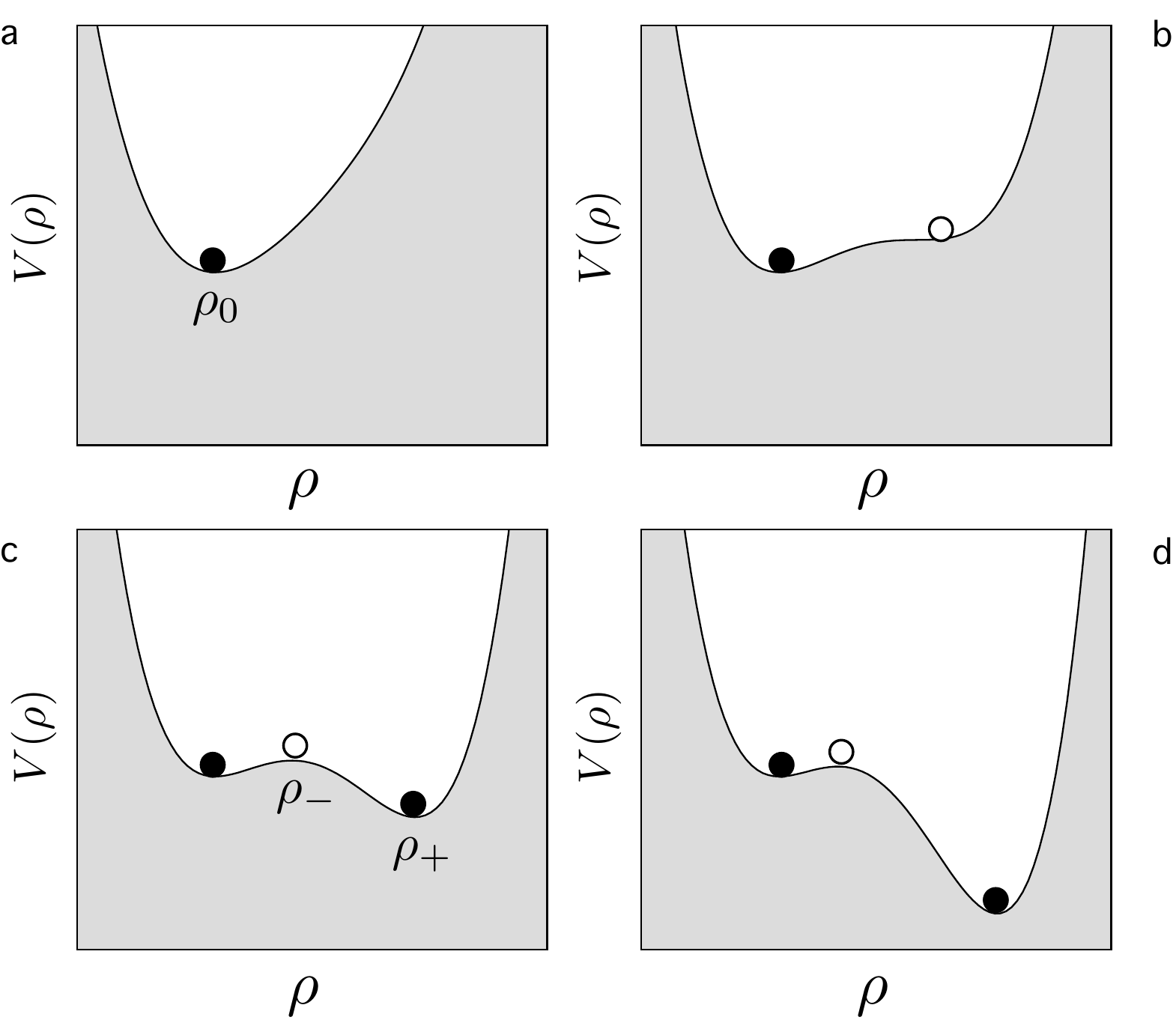}

          \caption{The potential function associated to the mean field theory for 
          the symmetric hypercycle, as given by \eqref{eq:cpcmft}. Here we fix $K=1$ and 
          $\delta=0.125$ which gives a critical point at $\Gamma_c=1$ (as in figure 1). 
          The parameter $\Gamma$ takes different values: (a) $\Gamma=0.5$, (b) $\Gamma=\Gamma_c=1.0$, 
          (c) $\Gamma=1.25$ and (d) $\Gamma=1.25$, respectively.}
        \label{fig:2}
        \end{center}
      \end{figure}

The marginal stability condition $\lambda_{\Gamma}=0$ provides 
the location of a first-order phase transition, as displayed in Fig.\ref{fig:2}. The 
trivial fixed point is stable under small perturbations, whereas $\rho_-$ and $\rho_+$ are 
unstable and stable, respectively. But these two nontrivial points only exist for $\Gamma > \Gamma_c$, and 
they collide at $\Gamma_c$ so that $\rho_1(\Gamma_c)=\rho(\Gamma_c)=K/4$. Here, a sharp transition governed by a {\it{saddle-node}} bifurcation takes place. As expected, if degradation is negligible ($\delta\to 0$) then $\rho_+(0)\to K/2$.

An alternative approach to stability is obtained by assuming that the dynamics is derived from a potential function, 
which allows us to write our system as
\begin{equation} 
{d \rho \over dt}= - {\partial V(\rho) \over \partial \rho} ,
\end{equation}
which in our case reads
\begin{equation} 
V(\rho) = \frac{\delta}{2} \rho^2 - \frac{\Gamma}{3} \rho^3+\frac{\Gamma}{2}\rho^4 .
\end{equation}

This function is such that its minima and maxima correspond to the 
stable and unstable fixed points, respectively. Four snapshots of this landscape are displayed 
in Fig.\ref{fig:2}. It can be noticed that the single-well potential is deformed as we approach $\Gamma_c$ and two 
minima appear after we cross it, with the valley associated to the stable hypercycle becoming 
deeper for larger $\Gamma$, while $\rho_0$ becomes a metastable state of the system \cite{Strogatz2014, RVSolePHT}. 

\section{Field theory}

In this section we approach the problem of stochastic hypercycles along the lines of previous 
work involving a field theory analysis of nonequilibrium systems, particularly for 
reaction-diffusion chemical systems \cite{Peliti85, Doi76a, vanWij01, Hin2000, Tau2005,Tau2014, Kay, Mattis1998}.   
In order to address the stochastic, spatial behavior of our two-member hypercycle, we consider a spatially-extended system that is discretized as a lattice in $d$ dimensions. 
Our goal is to define the conditions under which the hypercycle is expected to 
be stable and avoid the absorbing state and determine the role of dimensionality.

\subsection{Reaction-diffusion model}

Let us consider an  $L^d$ lattice of cell size $l$. In each lattice site we allow the system to evolve under the 
second-order, catalytic reactions
\begin{equation}
\label{eq:r1}
A  + B \buildrel \mu \over  \longrightarrow (2A + B, A +2B)
\end{equation}
as well as the (linear) decay transitions
\begin{equation}
\label{eq:r2}
(A, B) \buildrel  \delta  \over  \longrightarrow \varnothing .
\end{equation}
Here, $\mu$ stands for the reaction rate of the catalytic processes of $A$ via presence of $B$ and viceversa. We also let both species undergo diffusion at equal rate $D$. In addition to the reactions above, one needs to introduce extra processes that render the hard-core repulsive forces \cite{Tau2014},
\begin{equation}
\label{eq:r3}
(2A  + B,\,A+2B) \buildrel \lambda \over  \longrightarrow (2A,2B)
\end{equation}
and, similarly, we have 
\begin{equation}
\label{eq:r4}
(2A  + B,\,A+2B) \buildrel  \lambda^{\prime} \over  \longrightarrow A  + B  .
\end{equation}
This can be shown to be equivalent to van Wijland's construction \cite{vanWij01} for hard-core diffusive particles. Later, we will set $\lambda=\lambda^{\prime}$ for simplicity. Now the potential function becomes bounded from below and it is possible to recover the first order phase transition predicted in the previous section (see Fig.\ref{fig:2}). 
Spatial dynamics can be implemented via a master equation for all lattice sites $r,s\in\{ 1,\ldots,(L/l)^d \}$

\begin{widetext}
\begin{eqnarray}
\label{eq:master}
\frac{\partial}{\partial t} P(\vec{n},\vec {m};t) &=& \sum\limits_{\{r,s\}} D[(n_s+1)P(\ldots,n_r-1,n_s+1,\ldots,\vec{m};t)-n_rP(\vec{n},\vec {m};t)]+ \nonumber\\
&&\sum\limits_{r} \delta[(n_r+1)P(\ldots,n_r+1,\ldots,\vec{m};t)-n_rP(\vec{n},\vec{m};t)]+ \nonumber\\
&&\sum\limits_{r} \mu[(n_r-1)m_rP(\ldots,n_r-1,\ldots,\vec{m};t)-n_rm_rP(\vec{n},\vec {m};t)] +\\
&&\sum\limits_{r} \lambda[n_r(n_r-1)(m_r+1)P(\vec{n},\ldots,m_r+1,\ldots;t)-n_r(n_r-1)m_rP(\vec{n},\vec{m};t)]+ \nonumber\\
&&\sum\limits_{r} \lambda^{\prime}[(n_r+1)n_r(m_r+1)P(\ldots,n_r+1,\ldots,m_r+1,\ldots;t)-n_r(n_r-1)m_rP(\vec{n},\vec{m};t)] + \{\vec{n}\leftrightarrow \vec{m} \,, \nonumber \}
\end{eqnarray}
\end{widetext}
\noindent
where $D$ is the diffusion coefficient, $r$ is summed over all the lattice sites, and $s$ over the
nearest neigbors of $r$. The first line in the RHS of
\eqref{eq:master} implements diffusion through a random hopping of
particles between nearest neighbor sites.  The rest of the lines generate 
the interactions from \eqref{eq:r1}-\eqref{eq:r4}. Initial conditions, $P(\vec{n} ,\vec{m}, 0),$ are typically chosen as a Poisson distribution, with an average
density per site equal for both types of particles.

\subsection{Doi-Peliti second quantisation approach}

Following the Doi-Peliti second quantisation procedure \cite{Tau2005, Tau2014, Kay}, we 
construct a field theory for two species of diffusive molecules $A$ and $B$ 
and write down an Euclidean action upon which we will be able to apply perturbation theory.

We introduce two sets of creation
and anihilation operators at each lattice site, 
\begin{eqnarray}
&&\aA{r}\state{n_r}=n\state{n_r-1}\, \, \,\, \, \bA{r}\state{m_r}=m\state{m_r-1} \nonumber\\
&&\aC{r}\state{n_r}=\state{n_r-1} \, \, \,\, \,\, \, \, \, \bC{r}\state{m_r}=\state{m_r-1} \nonumber
\end{eqnarray}
which fulfill
the standard commutation rules 
\begin{equation}
  \Big{[} \aA{r}, \aC{s} \Big{]} = \left[ \bA{r}, \bC{s} \right] = \delta_{rs}.
\end{equation}
By incorporating these rules, the operators have a bosonic character. 
Using the vacuum state $\state{0}$, defined by $$\aA{r}\state{0} =
\bA{r} \state{0} =0$$ we construct an orthonormal basis of states
$\state{\vec{n},\vec{m}}$ for the Fock space, defined by

\begin{equation}
  \state{\vec{n},\vec{m}} = \prod_r (\aC{r})^{n_r} (\bC{r})^{m_r} \state{0},
\end{equation}

In terms of this
Fock space, the state of the system is given by the vector state $\state{P(t)}$, defined as

\begin{equation}
  \state{P(t)} = \sum_{\vec{n},\vec{m}} P(\vec{n} ,\vec{m},
  t)\state{\vec{n},\vec{m}}. 
\end{equation}
In terms of this vector state, the master equation 
is analogous to a Schr\"odinger equation in imaginary time, namely
\begin{equation}
  \label{eq:schro}
  \frac{\partial}{\partial t} \state{P(t)} = -\hat{H} \state{P(t)},
\end{equation}
with a Hamiltonian $\hat{H}=\hat{H}_{D}+\hat{H}_{\delta}+\hat{H}_{\mu}+\hat{H}_{\lambda}+\hat{H}_{\lambda^{\prime}}$ defined by diffusion terms plus interactive terms
\begin{eqnarray}
  \label{eq:hamil}
  && \hat{H}_{D} =  \frac{D}{l^2}  \sum_{\left< r s \right>} \left\{(\aC{r} -
    \aC{s})(\aA{r} - \aA{s}) + (\bC{r} -\bC{s})(\bA{r} -
    \bA{s}) \right\} \nonumber \\
  && \hat{H}_{\delta} = \delta \sum_{r} \left\{  (\aC{r} -1)\aA{r} + (\bC{r} -1)\bA{r} \right\} \nonumber\\
  && \hat{H}_{\mu} = \mu \sum_{r}  (\aC{r}+\bC{r}-2) \aC{r} \bC{r} \aA{r} \bA{r} \,.
\end{eqnarray}
For the sake of simplicity, we skip $\hat{H}_{\lambda}$ and $\hat{H}_{\lambda^{\prime}}$ terms. Eq.~\equ{eq:schro} can be formally solved in terms of the operator $\hat{H}$
yielding 
\begin{equation}
  \state{P(t)} = \exp ( -  \hat{H} t )
  \state{P(0)}.
  \label{eq:solutionop}
\end{equation}
This defines a Schr\"odinger equation that is real and thus the approach departs from the standard 
many-body quantum mechanics. In particular, the states are linear functions of the probabilities, instead 
of the amplitudes of probability.

     \begin{figure*}
        \begin{center}
          \includegraphics[width= 13 cm]{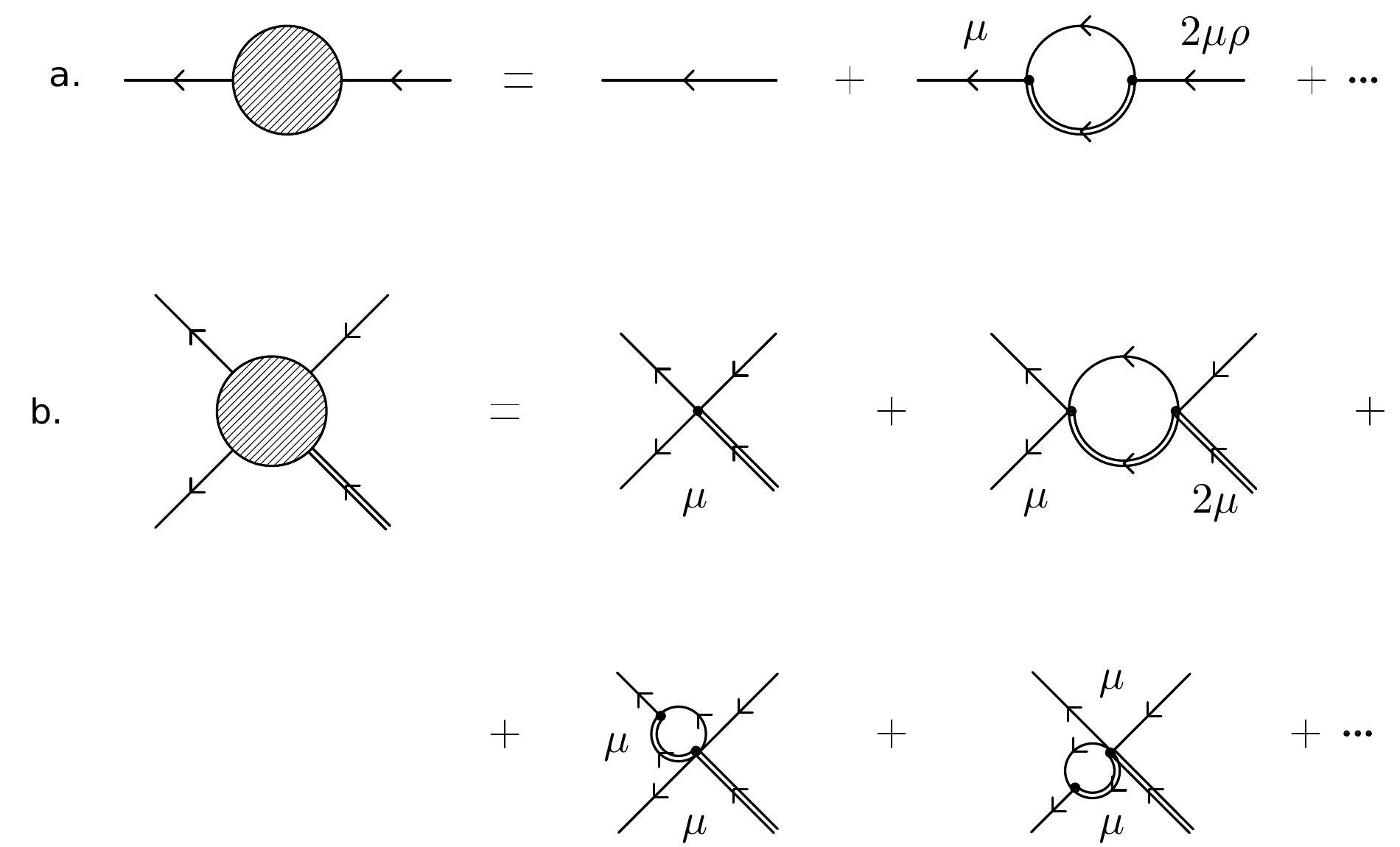}
          \caption{Computation  shown in (a) is associated to the propagator renormalization. Which couples to the mass (decay) coupling renormalization. On the other hand, (b) is the $\Gamma_{2,2}$ vertex computation, that allows the one-loop renormalization of the coupling $\mu$. Time flows to the left. Simple lines stand for the gaussian propagator of $a$-fields, $G_0^a$ and double lines for $b$-fields, $G_0^b$.}
          \label{fig:3}
        \end{center}
      \end{figure*}

\subsection{The Hypercycle action}

From the previous discrete set up, it is posible to realize a continuum limit considering that the scale at which
we are interested to perform measures is much larger than the size of the lattice cells. In fact, this is equivalent 
to introducing a physical cutoff $\Lambda\sim\frac{1}{l}$. 

In order to achieve a field theoretical description, a coherent state representation is introduced by setting 
\begin{eqnarray}
\aA{r}\state{\phi_r}=\phi_r\state{\phi_r}\, \, \,&& \, \, \, \, \bA{r}\state{\psi_r}=\psi_r\state{\psi_r} \nonumber\\
\astate{\phi_r}\aC{r}=\astate{\phi_r}\phi_r^* \, \, \,&& \, \,\, \astate{\psi_r}\bC{r}=\astate{\psi_r}\psi_r^* \ .\nonumber
\end{eqnarray}

This is chosen such that the scalar product is $ \big{<} \phi_r^*  \state{\phi_r}=e^{|\phi_r|^2}$, this structure will allow a resolution of unity \cite{Kay} so that it will be possible to treat $\phi_r^*$ and $\psi_r^*$ as auxiliary fields independent from $\phi_r$ and $\psi_r$, respectively. Thus, after taking the continuum limit,

\begin{eqnarray}
\{\phi_r\}\rightarrow\phi(\vec{x},t)\, \, \,&& \, \, \, \, \{\psi_r\}\rightarrow\psi(\vec{x},t) \nonumber\\
\{\phi_r^*\}\rightarrow\tph(\vec{x},t)\, \, \,&& \, \,\, \{\psi_r^*\}\rightarrow\tps(\vec{x},t) \, ,\nonumber
\end{eqnarray}
the system is mapped into a statistical field theory \cite{Tau2005}. Hence, the action describing the two-member reaction-diffusion hypercycle is given by

\begin{eqnarray}{\label{action}}
S[\tph,\phi,\tps,\psi] = \int d^dx\int dt  \Big\{ \tph\partial_t\phi+ \tps\partial_t \psi+\sum_{i=0}^3\H_i\Big\}_. 
\end{eqnarray}

The interaction parts $\H_i$, $i\in\{0,\ldots,3\}$ correspond to the reactions \eqref{eq:r1}-\eqref{eq:r4}, respectively. The MF solution is recovered when one-point expected values are taken and fluctuations are ignored. Thus, we read off the associated vacuum expected values (vev) for each field: $\langle \tph \rangle = \langle \tps \rangle = 1$, while, at the active phase, $\langle \phi \rangle = \langle \psi \rangle = \rho_+$. 

Since we are interested in probing the system fluctuations around the active stable solution (see Fig.\ref{fig:2}c), all fields need to be shifted to their corresponding minimum before implementing weak coupling perturbative RG methods:

\begin{eqnarray}
& \begin{cases} \tph\rightarrow\bar{a}+1 \\ 
       \tps\rightarrow\bar{b}+1\\
\end{cases}
& \begin{cases} \phi\rightarrow a+\rho_+ \\ 
       \psi\rightarrow b+\rho_+ \ . \nonumber\\
\end{cases}
\end{eqnarray}

This shift leads us to the a modified field theory action \eqref{action} with $a, \ba, b, \bb$ fields \cite{Initcond}. Now, the interaction terms obtained are

\begin{eqnarray}
\H_0=-D\big(\ba\nabla^2a + \bb\nabla^2b\big)+ \delta\big(\ba a+\bb b\big) +\delta\rho\big(\ba+\bb\big)\ \ \ \ \ \ \ \ \ \  &&\nonumber\\
\H_1= -\mu(\ba+\bb)(\ba+1)(\bb+1)(a+\rho)(b+\rho) \ \ \ \ \ \ \ \ \ \ \ \ \ \ \ \ \ \ \  && \nonumber\\
\H_2= \lambda\Big{[}\bb(\ba+1)^2(a+\rho)+\ba(\bb+1)^2(b+\rho)\Big{]}(a+\rho)(b+\rho) && \nonumber\\
\H_3= \lambda^{\prime}\Big{[}\ba(a+\rho)+\bb(b+\rho)\Big{]}(\ba+1)(\bb+1)(a+\rho)(b+\rho) . \  && \nonumber
\end{eqnarray}

Here onwards $\rho_+$ is replaced by $\rho$ for simplicity. Computing the classical solution by functional analysis on the system's action, i.e., 
$$\left ( \frac{\delta S}{\delta \ba} \right)_{\ba=0} = \left ( \frac{\delta S}{\delta \bb}\right)_{\bb=0}=0 \ ,$$
and ignoring all fluctuations ($n$-point correlation functions), we recover the MFT equations \eqref{MFT1} and \eqref{MFT2} with $\lambda=\lambda^{\prime}=K^{-1}$.

\subsection{Perturbative expansion and RG flow}

Firstly, let us suppose that the steady microscopic configuration is such that the system sits somewhere over the tipping point of Fig. \ref{fig:2}. Thus, we would like to compute how fluctuations can drive the system down to its inactive absorbing phase and figure out under which conditions the stationary state of the system will remain active as we let the scale flow towards the infra-red (IR) limit.

Via straightforward power counting we can identify that there is a dimension below which $\H_1$ operators become marginally relevant. Introducing a diffusive temporal scaling $[t]=\kappa^{-2}$, where $\kappa$ is the momentum scale, and taking $D$ as an adimensional constant we have
\begin{eqnarray}
&& [a]=[b]=[\rho]=\kappa^d \ , \ \ \ \ \ [\ba]=[\bb]=\kappa^0 \nonumber \\ 
&& [\delta]=\kappa^2 \ , \ \ [\mu]=\kappa^{2-d},\ \ [\lambda]=\kappa^{2-2d} \ .\nonumber
\end{eqnarray}

\begin{figure}
        \begin{center}
          \includegraphics[width= 8.2 cm]{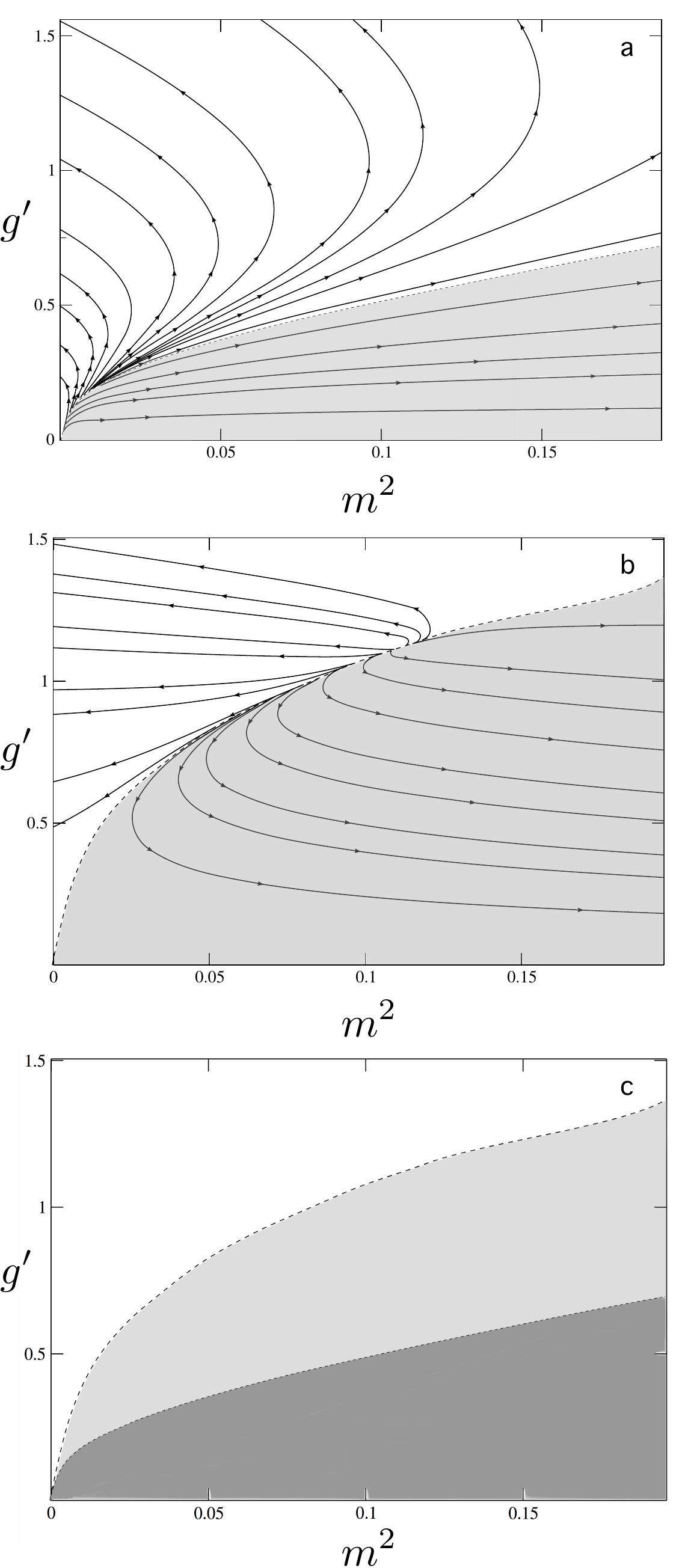}

          \caption{Coarse-graining scale flow towards the IR ($b\to\infty)$ of the couplings for the dimensionality $d\leq2$ (a) and $d=3$ (b). The shaded areas correspond to an inactive phase on the large-scale limit. Figure (c) corresponds to a superposition of (a) and (b). Here a drastic shift is observed as we let the system undergo rescaling from a 2D setup to a 3D one. In particular, mortality becomes predominant as $d$ is increased. The numerical analysis has been carried out using an RK4 algorithm. For numerical simplicity, on the y-axis $g^{\prime}\equiv(2/\pi)g$ is plotted instead of $g$. Since $n\geq1/4$, figures (a) and (b) are obtained for $n=1/4$.}
        \label{fig:RGFlow}
        \end{center}
      \end{figure}

Henceforth, for $d=2$ the coupling constant $\mu$ becomes marginal. Note, that the system is not posed near a critical state, thus, we cannot extract universal properties from a perturbative field theoretical computation. Nonetheless, it is possible to implement a coarse-graining process at the vecinity of $d=2$, i.e., $\epsilon=2-d$ and work up to one-loop expansion using the Feynman diagrams shown in Fig.\ref{fig:3}. This scheme will ultimately lead to the construction of a phase diagram (Fig.\ref{fig:RGFlow}c).

Let us note that for the shifted action \eqref{action} a new mass term is obtained, namely, $M^2:=\delta-\mu\rho+3\lambda\rho^2$ \cite{positivmass}. This mass coupling plays the role of an effective decay rate for the active system and it lays on the denominator of the fields propagators.
From the shifted action \eqref{action} we derive an effective action valid only in the vicinity of $d=2$ by only inculding $\H_0$ and $\H_1$ interactions. A total of 26 vertices are obtained. After introducing $\Lambda$ as the large momentum cutoff, i.e., $\Lambda\sim 1/l$, we 
define the adimensional couplings 

\begin{eqnarray}
m^2\equiv\Lambda^{-2}M^2/D \ ,\nonumber\\
n\equiv\Lambda^{-d}\rho \ ,\nonumber\\
g\equiv\Lambda^{-\epsilon} \mu/D \ .\nonumber
\end{eqnarray}

The one-loop perturbative computation is derived from the vertex-functions $\Gamma_{1,1}$ and $\Gamma_{2,2}$ which diagramatic expansions are shown in Fig. \ref{fig:3}. These calculations generate coarse-grained couplings $g_R$ and $m^2_R$. Upon differentiation with respect to the momentum scale $\kappa$ the couplings flows are obtained. 

Next, we introduce a length-scale parameter $b=\log(s)$ with $s=l^{\prime}/l$, where $l^{\prime}$ is the coarse-grained scale. Thus, $b\sim\kappa^{-1}$, and $b\to\infty$ will correspond to the IR limit. The results are given by:

\begin{eqnarray}
&& \frac{dm^2}{db}=2m^2-\frac{n}{\pi}\frac{g^2}{1+m^2} \ ,{\label{RGflow}} \\
&& \frac{dg}{db}=\epsilon g+\frac{2}{\pi}\frac{g^2}{1+m^2}\ . {\label{RGflow2}}
\end{eqnarray}

The analysis of these flow equations allows us to probe the dynamics of the system at different length scales. Lines in Fig.\ref{fig:RGFlow} are solutions to these equations given different initial conditions, which are understood as the microscopical coupling constant values. As $b$ grows the couplings are readjusted to a new value that renders the dynamics of the new and larger length scale. From these results we read how the system's activity pervades under rescaling.

The sign of the $\epsilon$ term in \eqref{RGflow2} shifts the nature of the trivial fix point from unstable in $d\leq2$ (Fig.\ref{fig:3}a) to $d=3$ (Fig.\ref{fig:3}b) for the coupling $g$.

Solutions plotted in Fig.\ref{fig:3} are restricted from a perturbative point of view \cite{perturbative}, thus, the {\it{active/inactive}} phase diagram provided by Fig.\ref{fig:RGFlow}c is not exact once a certain threshold is exceeded. Nonetheless, the criterion used to select the curves that drive the system into extinction is a conservative one. If the mass coupling flows above the values shown in Fig. 4a and 4b, then we assume that the system falls into extinction, eventhough at the strict $b\to\infty$ limit that curve might return to a zero mass value.

In this approach, a radical transition in the shape of the active/inactive curve shown in Fig.\ref{fig:3}c is promptly observed when allowing the dimensionality go from 2D to 3D, suggesting a clear advantage for the system's survival on a 2-dimensional layout. This observation is a direct result of the field-theoretical formalism. This is the main result of our paper, indicating that the survival of a cooperative molecular 
system (as the one described here) is strongly enhanced under a dynamics constrained on a surface. This can be easily interpreted 
in terms of the well known conjecture that reaction kinetics might have been strongly facilitated in two-dimensional substrates 
where limited movement of molecules is involved. This would not be the case in a three-dimensional context, where 
a high effective probability of extinction is expected.

\section{N-member homogeneous hypercycle}

In this section we will briefly outline how dimensionality affects a generalization of hypercycles known as the 
N-member perfectly cyclic systems \cite{Eigen} using the same field-theoretical setup as in the previous analysis. We have already defined 
in (3) the basic deterministic equations for such a general hypercycle. Extensive work has shown that complex dynamics 
can emerge, including oscillations and chaos \cite{Eigen, FontanariPRE2002, McKaskill1997, Hofbauer}. Is it possible to generalise our previous 
approach to an arbitrary hypercycle? In this section we aim to address this problem using the field theory formalism. 

\begin{figure}
        \begin{center}
          \includegraphics[width= 5.5 cm]{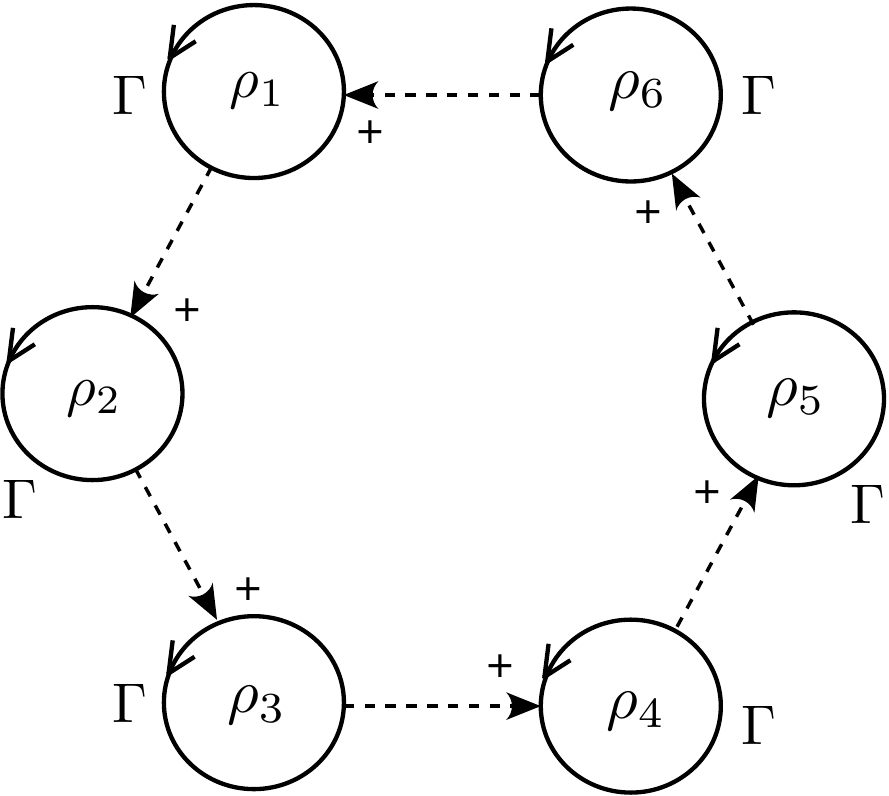}

          \caption{Higher-order hypercycles can be build by the cooperation of multiple 
          replicators forming a closed loop. Here we show a homogeneous $N=6$ hypercycle, 
         where each member helps the next one. Decay transitions are not displayed.}
        \label{fig:NHYP}
        \end{center}
       \end{figure}

Consider a perfectly ordered symmetric hypercycle with pair-connected species, that is,
\begin{eqnarray}
& \begin{cases} I_k\buildrel \delta \over \longrightarrow\varnothing \\ 

       I_k+I_{k+1}\buildrel \Gamma \over  \longrightarrow I_k+2I_{k+1}\\
\end{cases}
\end{eqnarray}
with $k=1,\ldots,N-1$, plus closing loop reactions $I_N\buildrel \delta \over \longrightarrow\varnothing$ and $I_N+I_1 \buildrel \mu \over  \longrightarrow I_N+2I_1$ (see Fig.\ref{fig:NHYP}). 

In this case, the extra hard-core repulsive reactions that need to be imposed are provided 
by the reactions:
\begin{eqnarray}
& \begin{cases} (2I_i+I_j,I_i+2I_j)\buildrel \lambda \over \longrightarrow I_i+I_j \nonumber \\ 

       (2I_i+I_j,I_i+2I_j)\buildrel \lambda^{\prime} \over  \longrightarrow (2I_i,2I_j) \nonumber \\
       I_i+I_j+I_k \buildrel \lambda^{\prime\prime} \over \longrightarrow I_i+I_j \nonumber \\
\end{cases}
\mbox{\rm for\,} i\neq j \neq k \nonumber
\end{eqnarray}

Upon application of Doi-Peliti formalism, the $\H_0$ and $\H_1$ contributions are obtained:

\begin{eqnarray}
\label{eq:Nham}
\H_0&=&-D\sum_{k=1}^{N}\bar{\phi}_k\nabla^2\phi_k+ \delta\sum_{k=1}^{N}\bar{\phi}_k\phi_k \nonumber \\
\H_1&=&-\Gamma\sum_{k=1}^{N-1}\bar{\phi}_{k+1}(\bar{\phi}_{k+1}+1)(\bar{\phi}_k+1)\phi_{k+1}\phi_k  \nonumber \\
&&\, + \,\, \bar{\phi}_{1}(\bar{\phi}_{1}+1)(\bar{\phi}_N+1)\phi_{1}\phi_N , 
\end{eqnarray}
note that the $\phi_k$ fields here are not shifted to their non-trivial vacuum expected values.

From the formal linear structure provided by the hamiltonian dynamics above, it can be argued that following the previous steps leads to a similar conclusion regarding the role of dimensionality for the N-member hypercycles. However, it is crucial to point out that, even though we can probe the system using a coarse-graining scheme as before, both initial conditions and intermediate configurations impose a major constraint for this kind of systems. This can be understood via two different mechanisms. On the one hand, it is well known  that, for $N>4$, deterministic dynamics of coexistence in hypercycles is given by oscillatory behaviour \cite{Hofbauer,Silvestre,PRECamposFontanari}. These oscillations drive the system close to its extinction state, whence, intrinsic-external noise or spatially induced fluctuations may force the system into its inactive phase. 

Even though trying to compute statistical measures using \eqref{eq:Nham} is a task beyond the scope of this article, it can be conjectured that, provided that sufficiently good initial conditions are given, then {\it{surface dynamics}} stands as an evolutionary selective advantage for these type of hypercyclic systems in contrast with higher dimensional layouts.

\section{Conclusions}

Several studies concerning the origins of life and cooperation among molecular replicators have 
approached the problem using a physics-like perspective \cite{Dyson1982,Kauffman1993,Ferreira2002,Goldenfeld2011,Wu-Higgs2012}. 
By reducing the system's complexity to its minimal logic rules, we can search for unifying principles of 
emergent dynamics and universality. 

In order to overcome the limitations imposed by error-prone replication in a prebiotic scenario, 
it has been suggested that a cooperative loop of coupled reactions, the hypercycle, is a necessary 
condition to achieve higher levels of complexity \cite{Eigen,SchusterEvo,Smith97}. Similarly, 
cooperative dynamics might be a requirement to sustain complex ecosystems \cite{Wilkinson2006}. The role played by 
spatial constraints as well as stochastic effects has been previously considered by a number of authors \cite{SolomonPNAS2000}, 
indicating that it can play a key role in making possible the stabilisation of hypercycles, particularly 
in relation with their resilience against parasites and other sources of disturbance. These studies have indicated 
that in two dimensions the hypercycle can be more robust, developing spatial structures that protect them (in particular) 
parasites. It is worth mentioning that the study of deterministic hypercycles in three dimensions (using 
reaction-diffusion equations) reveals that structures become unstable thus making the resulting 
dynamics less likely to persist \cite{3dhyper1996}.

In this paper we have analysed a spatially-extended, reaction-diffusion field theory of hypercyclic replicators. 
The main target of our study is the simplest, symmetric two-member hypercycle involving two cooperative partners. 
The model includes both second-order reactions based on interactions among components of each class as well 
as a linear degradation term. We have studied the general conditions for hypercycle survival as provided by the 
RG flow equations. It has been shown that the hypercycle has robust properties (escaping from extinction) for $d_c=2$
dimensions, whereas for $d > d_c$ a dramatic change happens, with a much less 
reduced survival of the molecular cooperators. These results support the view that early life 
required a surface context to persist and evolve complexity. Further analysis based on field theoretical 
approaches should consider the role of asymmetries in the hypercycle \cite{AsymmetricHypercycle} as well as an explicit 
consideration of parasites. Similarly, other classes of replication dynamics (based on templates) as 
well as additional components such as error tails \cite{errorprone} will be explored elsewhere.

\begin{acknowledgments}
\small{The authors thank J. Sardany\'es and G. Torrents for helpful discussions. This work was supported by 
the Botin Foundation, by Banco Santander through its Santander Universities Global Division and by 
the Santa Fe Institute.}
\end{acknowledgments}


\begin{thebibliography}{99}
\bibitem{HYPalternative} 
This assumption that both densities are equal is essentially correct for an infinite system at the 
steady state. A different derivation of a symmetric system can be obtained using the standard 
constant population constraint \cite{Eigen} where the equations for the hypercycle are now written as 
$d\rho_i/dt=\Gamma_{ij}\rho_1 \rho_2 - \delta_i \rho_i - \rho_i \Omega(\rho_1,\rho_2)$ with 
$i=1,2$. The last term in these equations introduces an effective outflow, and 
using the CPC means that $\rho_1+\rho_2=1$ and this leads, for the symmetric case, 
 to $\Omega(\rho_1,\rho_2)=2\Gamma \rho_1 \rho_2 - \delta$. The dynamics is thus described by a 
 different mean field equation, namely $d\rho/dt=\Gamma \rho(1-\rho)(1-2\rho)$ with two unstable states 
 $\rho=0,1$ and a stable coexistence point $\rho=1/2$, so no absorbing phase associated to 
 the hypercycle extinction is present.
\bibitem{Initcond}
After shifting $\tph$ and $\tps$ fields, bare time derivatives appear in the action \eqref{action} $\partial_ta$ and $\partial_tb$. Since we presuppose a steady state and therefore take the integration limits $t\to\pm\infty$ these terms may be dropped.
\bibitem{positivmass}
A full analysis following van Wijland's approach on hard-core repulsive action (\cite{vanWij01}) shows that, for our system, $M^2=\delta-\mu\rho e^{-3v\rho}$, where $v$ here is the ocupation number per lattice site. In this paper, we are including the ignorance of all higher order terms inside $\lambda$.
\bibitem{perturbative}
Since we are operating only at one-loop expansion, curves that travel far from the perturbative regime cannot be trusted.
\end{thebibliography}
\end{document}